\documentclass[twocolumn,a4paper,showpacs,showkeywords,superscriptaddress]{revtex4-1}
\usepackage{CJK,hyperref,graphicx,rotating,subfigure,amsmath,amssymb,amsfonts,delarray,CJK}
\usepackage{geometry}
\begin{document}
\title{Quantum information analysis quantum phase transitions in a one-dimensional $V_1$-$V_2$ hard-core-boson model}
\author{Jie Ren}
\affiliation{Department of Physics and Jiangsu Laboratory of Advanced
Functional Material, Changshu Institute of Technology, Changshu 215500, China}
\author{Xuefen Xu}
\affiliation{School of Mathematics and Physics,
Jiangsu Teachers University of Technology, Jiangsu 213001, China}
\author{Liping Gu}
\author{Jialiang Li}
\affiliation{Department of Physics and Jiangsu Laboratory of Advanced
Functional Material, Changshu Institute of Technology, Changshu 215500, China}

\date{\today}

\begin{abstract}
The entanglement entropy and quantum fidelity in a hard-core-boson model with nearest- and next-nearest-neighbor interactions are studied numerically. By using exact diagonalization and the density matrix renormalization group, the effects of interactions on entanglement entropy and fidelity susceptibility in the model are investigated. We focus our attention on looking for three quantum phase transitions. It is found that the first derivative of the entanglement entropy can indicate all of three phase transitions, while the fidelity susceptibility cannot predict the transition from superfluid to bond-order in a finite-size system.
\end{abstract}
\maketitle

\section{introduction}

In condensed matter physics, quantum phase transitions, which happen at zero temperature, have received much attention. In a quantum phase transition the properties of a many-body system will change dramatically, when a controlling parameter changes across critical point~\cite{Sachdev}. The phase diagram in the one-dimensional spinless fermion model has been determinded by many researchers. It is well known that a transition from Luttinger liquid to charge-density-wave phase occurs when the repulsive interaction is twice the hopping interaction. However,
long-range Coulomb interactions sometimes also play an important role in frustrated systems due to their geometrical structure. For example, a spinless fermion, including the nearest-neighbor and next-nearest-neighbor repulsive interactions can describe the low-energy electronic state of $CuO$ double chains \cite{Seo}, and the repulsions plays a significant role in the charge-ordered-insulator-metal transition.

The phases of spinless fermion including the nearest-neighbor as well as next-nearest-neighbor repulsive interactions are more complicated than one might expect. The model can be described by the following Hamiltonian:
\begin{eqnarray}
\label{eq1}
&H=
\displaystyle{\sum_i}[-t(c_i^\dagger c_{i+1}+h.c.)+V_1(c_i^\dagger c_i-\dfrac{ 1}{ 2})(c_{i+1}^\dagger c_{i+1}-\dfrac{ 1}{ 2}) \nonumber\\
&+V_2(c_i^\dagger c_i-\dfrac{1}{ 2})(c_{i+2}^\dagger c_{i+2}-\dfrac{1}{ 2})],
\end{eqnarray}
where $c^\dagger_i(c_i)$ is the fermionic creation (annihilation) operator at site $i$, $t$ is the hopping amplitude, and $V_1$ and $V_2$ are the nearest-neighbor and next-nearest-neighbor Coulomb interactions. The fermionic Hamiltonian (\ref{eq1}) can be mapped onto the identical Hamiltonian for hardcore bosons, which is given by
\begin{eqnarray}
\label{eq2}
&H=
\displaystyle{\sum_i}[-t(b_i^\dagger b_{i+1}+h.c.)+V_1(b_i^\dagger b_i-\dfrac{ 1}{ 2})(b_{i+1}^\dagger b_{i+1}-\dfrac{ 1}{ 2}) \nonumber\\
&+V_2(b_i^\dagger b_i-\dfrac{1}{ 2})(b_{i+2}^\dagger b_{i+2}-\dfrac{1}{ 2})],
\end{eqnarray}
where $b^\dagger_i(b_i)$ is the hardcore boson creation (annihilation) operator at site $i$, and $t$ is the hopping integral and taken as the unit of energy in the paper. $V_1$ and $V_2$ are the nearest-neighbor and next-nearest-neighbor interactions. This mapping can be done by using a Jordan-Wigner transformation, as both models have the same spectrum. The phase diagram for hard-core bosons is relevant to the corresponding spinless-fermion model, as is shown in Refs. \cite{Duan,rigol}. It is noted that the Luttinger-liquid phase in the spinless-fermion model is relevant to the superfluid phase ({SF}) in the hard-core-boson model, and other phases have the same names in the hard-core-boson model and the spinless-fermion model. We also focus our analysis on the model (\ref{eq2}) at half filling. The Hamiltonian of the case $V_2=0$, can be solved exactly via Bethe ansatz. We find a transition at $V_1/t=2$ from the {SF} to the $(\cdots 01010101 \cdots)$ charge-density-wave ({CDW-I}) phase as $V_1$ is increased. When $V_2\neq 0$, the phase structure becomes rich. It is reported that there exist bond-order ({BO}) phase\cite{Schmitteckert,Duan,rigol}, which is somehow was missed in some research \cite{Zhuravlev,Poilblanc}. When $V_2$ is large, $(\cdots 00110011 \cdots)$ charge-density-wave phase({CDW-II}) with a four-site unit cell appears. Here, $1 (0)$ denotes the presence (absence) of a particle at a particular site.

Recently, due to the cross over field between quantum many-body theory and quantum-information theory, the ground-state entanglement entropy and fidelity have been used to qualify quantum phase transitions in the one-dimensional spin systems ~\cite{Amico,Gu, Abasto,Buonsante,You,Zhou01,Tzeng,Ren01,legeza}. Our goal is to combine these newly developed observables with accurate numerical calculations, and independently determine the phase diagram of the above model.

In the paper, we calculate the ground-state entanglement entropy and fidelity in the half-filled hardcore boson model including the nearest-neighbor as well as the next-nearest-neighbor repulsive interactions, and use them as good tools for searching for the phase transition points. The remainederof the paper is organized as following. The measurements and the details of the methods to obtain the ground state are introduced in Sec. II. The results for entanglement entropy are presented in Sec. III and for fidelity in Sec IV. Finally a discussion is given at last in Sec. V.

\section{measurements and methods}

The entanglement entropy  can be chosen as a
measurement of the bipartite entanglement, which can be used to detect
a quantum phase transition. The entropy is defined
as follows. Let $|g.s.\rangle$ be the ground state of a chain system, which can be divided into two parts A and B. The reduced density matrix of part A can be obtained by taking the partial trace over system B, which is given by
\begin{equation}
\label{eq3}\rho_{A}=Tr_{B}(|g.s.\rangle \langle g.s.|).
\end{equation}
The bipartite entanglement between parts
$A$ and $B$ can be measured via the entanglement entropy as

\begin{equation}
\label{eq4}S_{AB}=-Tr(\rho_{A}\log_2\rho_{A}).
\end{equation}

Another concept from quantum information theory, ground-state fidelity, can be applied to capture the existence of the quantum phase transitions. The general Hamiltonian of a quantum many-body system can be written as $ H(\lambda)=H_0+\lambda H_I $, where $H_I$ is the driving Hamiltonian and $\lambda$ denotes its strength. If $\rho(\lambda)$ represents a state of the system, the ground-state  fidelity between $\rho(\lambda)$ and $\rho(\lambda+\delta)$ can be defined as
\begin{equation}
\label{eq5}
F(\lambda,\delta)=Tr[\sqrt{\rho^{1/2}(\lambda)\rho(\lambda+\delta)\rho^{1/2}(\lambda)}].
\end{equation}
For a pure state $\rho=|\psi\rangle\langle \psi|$, Eq. \ref{eq5} can be rewritten as $ F(\lambda,\delta)=|\langle \psi(\lambda)|\psi(\lambda+\delta)\rangle|$. $F(\lambda,\delta)$ reaches its maximum value $F_{max}=1$ at $\delta=0$. By expanding the fidelity in powers of $\delta$,  and since the first derivative $\frac{\partial F(\lambda,\delta=0)}{\partial \delta}=0$, the fidelity can be written as
\begin{equation}
\label{eq6}
F(\lambda,\delta)\simeq1+\frac{\partial^2F(\lambda,\delta)}{2\partial\delta^2}\delta^2.
\end{equation}
Therefore, the average fidelity susceptibility $\chi(\lambda,\delta)$ can be given by~\cite{Buonsante}
\begin{equation}
\label{eq7}\chi(\lambda,\delta)=\lim_{\delta\rightarrow
0}\frac{2[1-F(\lambda,\delta)]}{N\delta^2},
\end{equation}
where $N$ is the length of the system.

As is well known, it is not easy to calculate the ground-state fidelity and entanglement entropy, because of the lack of knowledge of the ground-state wave function. The Hamiltonian (\ref{eq2}) is not exactly solvable except for $V_2=0$; we resort to exact diagonalization to obtain the ground state for small size, i.e., $N=20$. For large size, the method of the density-matrix renormalization-group(DMRG)~\cite{white,U} can be applied to obtain the ground state of model (\ref{eq2}). The total number of density matrix eigenstates held in a system block is $m=250$ in the basis truncation procedure. The Matlab code for the finite-size density-matrix renormalization group with double precision is used to evaluate the model with system sizes up to $N=100$, and the truncation error is smaller than $10^{-9}$. Both the above two methods are used to calculate the model (\ref{eq2}) with open boundary conditions.

\begin{figure}
\includegraphics[width=0.55\textwidth]{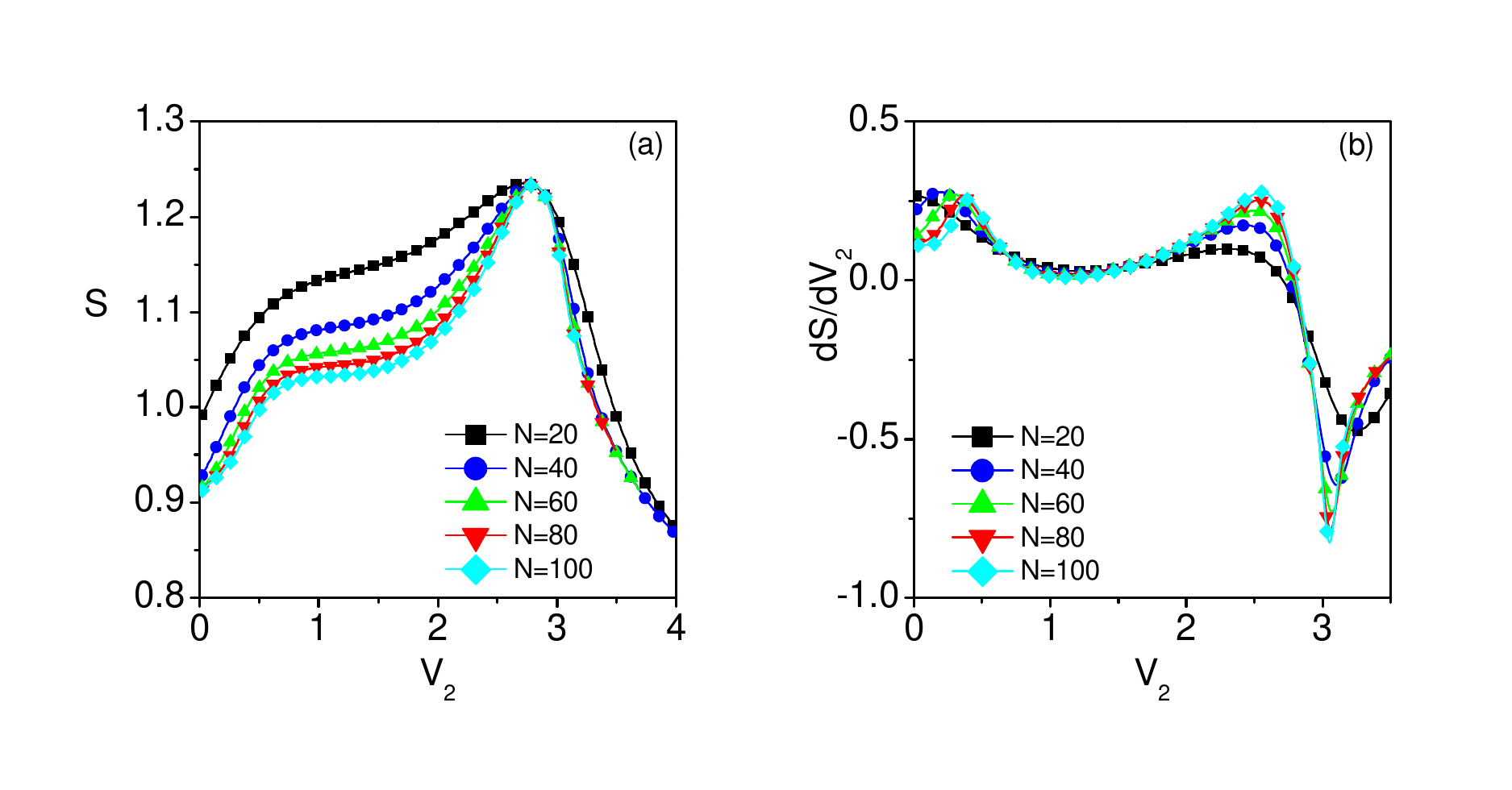}
\caption{\label{fig1} (Color online)  The  entanglement entropy of two central qubits (a) and its first derivative (b) are plotted as functions of $V_2$ for $V_1=4-V_2$ with $t=1$ for different system sizes. Here and in the following graphics all quantities are dimensionless. }
\end{figure}

\section{entanglement entropy}
\begin{figure*}
\includegraphics[width=0.650\textwidth]{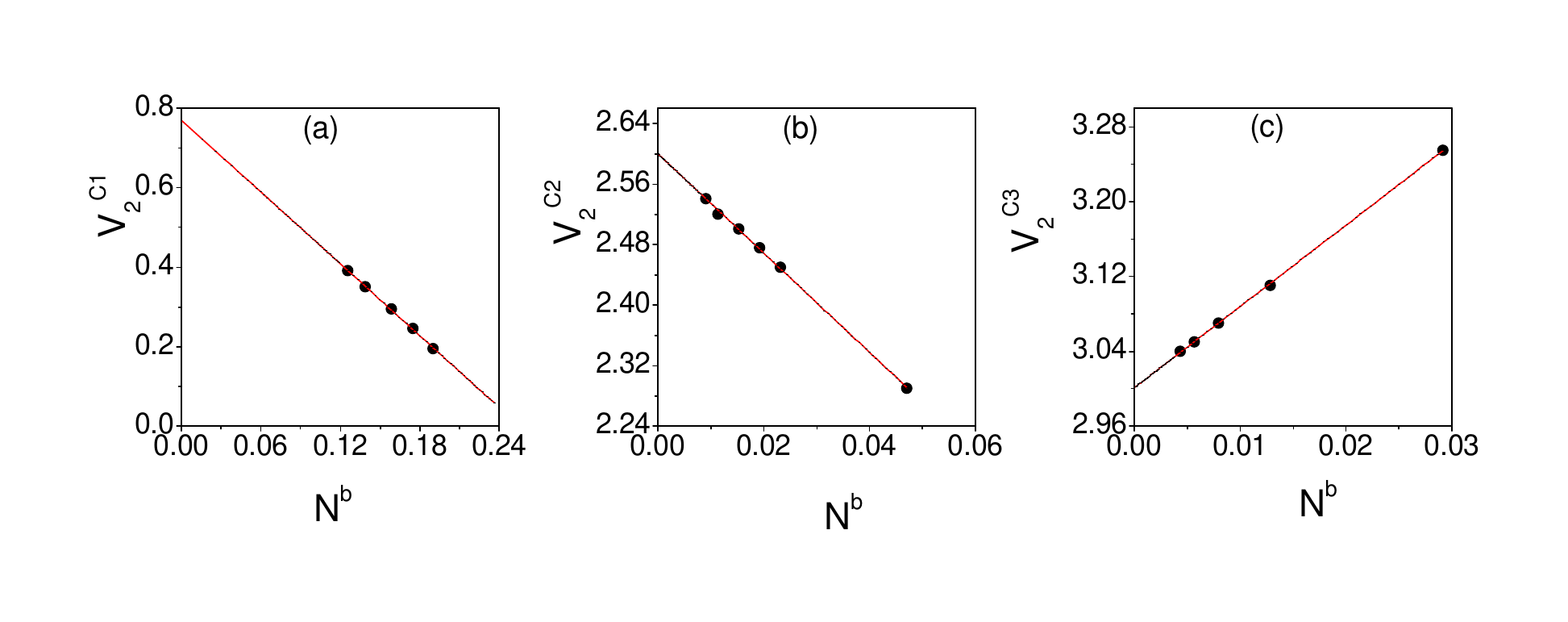}
\caption{\label{fig2} (Color online) The scaling behavior of the extreme points of $V_{2}$ versus $N^b$. The red lines are fixed lines.}
\end{figure*}

In order to avoid boundary effects, the entanglement entropy of two central two neighboring sites of a large system is calculated. By use of exact diagonalization and the DMRG, the entanglement entropy of the central two particles is plotted as a function of interaction for sizes $N = 20,40,60,80,100$
in Fig. \ref{fig1}(a), and the corresponding first derivatives are shown in Fig. \ref{fig1}(b). It is found that the entanglement entropy initially increases rapidly with increases of $V_2$. When  the interaction $V_2$ increases further, the entanglement entropy increases slowly, and reaches a peak value at $V_2=2.80$, which is independent of system size. As we know from Refs. \cite{rigol,Duan}, the peak of local entanglement does not coincide with the critical point. Considering the energy of the system in terms of the reduced density matrix of two spins at positions $i$ and $j$, the energy reads

\begin{equation}
\label{eq8}E_{i,j}=\sum_{i,j}Tr(H_{i,j}\rho_{i,j})
\end{equation}
where $H_{i,j}$  is the reduced Hamiltonian of the two sites at
positions $i$ and $j$, and the sum is over the total Hamiltonian of the system. $\rho_{i,j}$ is the reduced density matrix of two sites at positions $i$ and $j$. Moreover, the second-order derivative of
the energy is related to the derivative of the reduced density matrix. This is shown by

\begin{eqnarray}
\label{eq9}
\frac{\partial^2 E_{i,j}}{\partial V_2^2}=\sum_{i,j}[Tr(\frac{\partial^2 H_{i,j}}{\partial V_2^2}\rho_{i,j})+Tr(\frac{\partial H_{i,j}}{\partial V_2}\frac{\partial \rho_{i,j}}{\partial V_2})].
\end{eqnarray}
It is noted that a discontinuity in the second derivatives of the energy requires the divergence of at least one of the derivatives $\frac{\partial \rho_{i,j}}{\partial V_2}$ at the critical points \cite{wu}.The phase transition points sometimes are determined by the positions at which the first-order derivative of the two-site entanglement entropy takes on its maximal and minimal values \cite{Ren02,Liu}. Here, the CDW-II to BO transition is a second-order transition, so it should be captured by the first-order derivative of the two-site entanglement entropy. The phase transitions of {SF} to {BO} and {BO} to {CDW-II} are infinite-order Berezinskii-Kosterlitz-Thouless (BKT) transitions, which can also be predicted by the first-order derivative of the two-site entanglement entropy. The first derivative of the entanglement entropy is shown in Fig. \ref{fig1}(b). There are two peaks, whose locations move to high $V_2$ with increase in system size, and the location of the valley moves to lower $V_2$ with system size increases. These behaviors capture the {CDW-I} to {SF}, {SF} to {BO}, and {BO} to {CDW-II} phase transitions, respectively.

\begin{figure}
\includegraphics[width=0.250\textwidth]{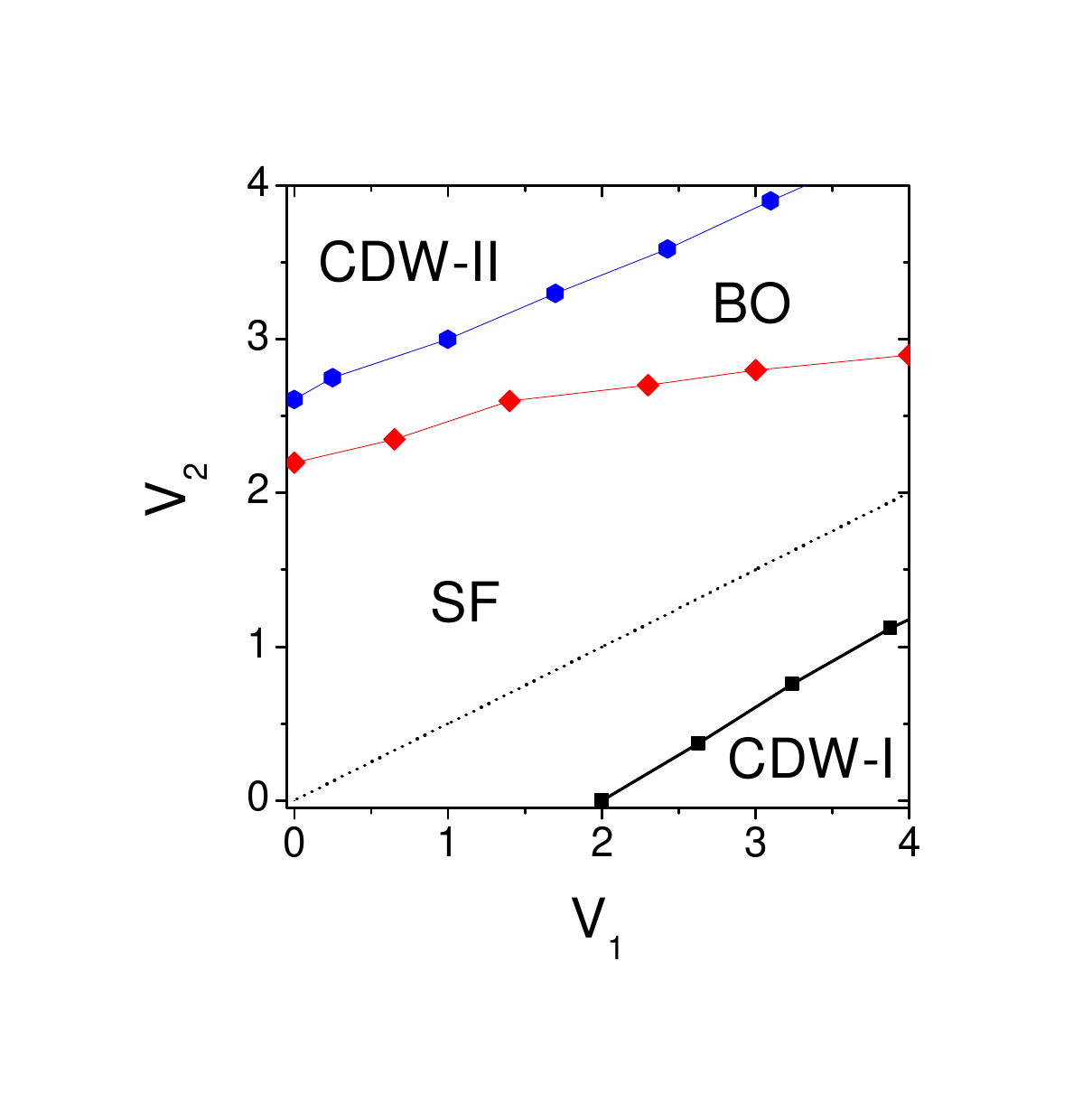}
\caption{\label{fig3} (Color online) The phase diagram of the model (\ref{eq2}) at half filling in the tilted $V_1$-$V_2$ plane. The dashed line corresponds to $V_2=V_1/2$, which is the boundary of the CDW-I and CDW-II phases when $t=0$. }
\end{figure}

The positions of the two peaks and one valley in the first derivative of the two-site entropy can be extrapolated to the thermodynamic limit. We investigate the size dependence of the positions of their extreme points. The changes of the critical points can be fitted by the formula $V_2 \sim V_{2}^c+aN^b$, where $a,b$ are constants and  $N$ is the size of the system. We plot the locations of the two peaks and one valley as functions of $N^b$. The fitting results are shown in Fig. \ref{fig2}. Here we obtain that $V_{2}^{c1}=0.77$, $a_1=3.012$, $b_1=-0.45$, $V_{2}^{c2}=2.60$, $a_2=-6,564$, $b_2=-1.02$, and  $V_{2}^{c3}=2.978$, $a_3=-0.9133$, $b_3=-0.621$.

In Fig. \ref{fig3}, we also show the phase diagram of the model, which is detected via the entanglement entropy. It is seen that our results for the CDW-I to SF boundary and the CDW-II to BO boundary are consistent with the results in Refs. \cite{rigol,Duan}.  The difference is in the BO to SF boundary.  The region of the BO phase is smaller than the region captured by the static structure factor. For $V_1=4$, we capture the BO to SF critical phase transition point $V_2^c=2.90$. It is consistent with the result obtained from the energy gap and larger than the result $V_2^c=2.55$ obtained from the static structure factor, which is also shown in Ref. \cite{rigol}.

\section{Fidelity}
\begin{figure}
\includegraphics[width=0.5\textwidth]{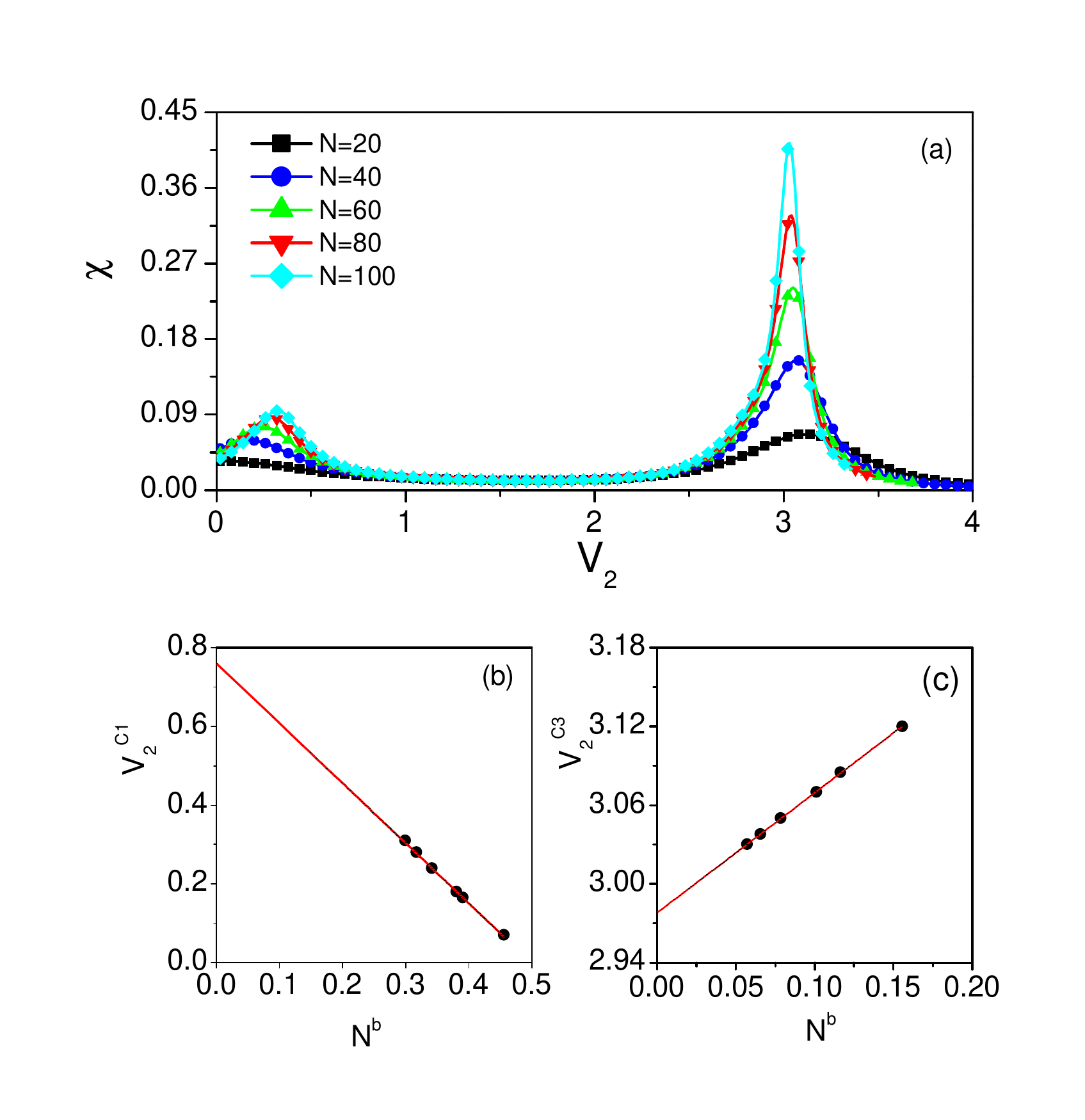}
\caption{\label{fig4} (Color online) (a) Fidelity susceptibility versus interaction $V_2$ for $V_1=4-V_2$ with $t=1$ for different system sizes. (b),(c) The scaling behavior of the extreme points of $V_{2}$ versus $N^b$. The red lines are fixed lines.}
\end{figure}

We apply the DMRG to calculate the ground-state fidelity susceptibility with system size $N$ up to $100$ and $\delta=0.01$. The ground-state fidelity susceptibility $\chi$ is plotted as a function of the interaction $V_2$ for different sizes in Fig. \ref{fig4} in dimensionless units. Two peaks in the fidelity susceptibility is found. The peaks of $\chi$ increase when the system size increases. One of the peaks' locations moves to high $V_2$ up to a particular value as the system size increases, and the other moves to a slightly lower $V_2$ down to a particular value as the system size increases. The fidelity measures the similarity between two states, while quantum phase transitions are intuitively accompanied by an abrupt change in the structure of the ground-state wave function. This primary observation motivates researchers to use the fidelity to predict quantum phase transitions.

A peak in the fidelity susceptibility indicates a quantum phase transition. Here, scaling of the extreme points of the fidelity susceptibility as the system length increases is also investigated. We find that the changes of the maximal points can also be fitted by the formula $V_2 \sim V_{2}^c+aN^b$, where $a,b$ are constants and $N$ is the number of the system. The results for the locations of the fidelity susceptibility can be used to investigate the quantum phase transition points in the thermodynamic limit. We plot the locations of the maximum fidelity susceptibility as a function of $N^b$ and show the numerical fit in Fig. \ref{fig4} (b) and \ref{fig4} (c). We obtain that $V_{2}^{c1}=0.76$, $a_1=-1.522$, $b_1=-0.262$, and  $V_{2}^{c3}=2.978$, $a_3=-0.9133$, $b_3=-0.621$.

The quantum phase transitions points are partially consistent with the results obtained from the entanglement entropy. It is shown that the fidelity susceptibility can predict the CDW-I to SF transition and the BO to CDW-II transition, however, it cannot predict SF to BO transition. As we know, the BO to CDW-II and SF to BO transitions are both infinite-order BKT transitions. The Numerical results clearly show that the BKT transition point is detectable to the BO to SF transition by divergence of the fidelity susceptibility \cite{Wang}. Our results are similar to the results, which the entanglement entropy predicts the BO phase in the extended Hubbard model \cite{Mund}, and the fidelity can not \cite{Li}. The BKT phase transition point sometimes cannot be well characterized by the ground-state fidelity for finite-size systems \cite{Chen}. A other reason may be that, according to the analysis of Ref. \cite{Cozzini}, a divergence in the fidelity susceptibility implies a quantum phase transition, but the converse is not true. This means that there are quantum phase transitions driven by particularly weak perturbations, where the fidelity susceptibility may not diverge. It would be interesting to use infinite-time-evolving block decimation\cite{vidal} to calculate the ground-state fidelity, and use it to detect the superfluid to bond order transition. Here we do not show the phase diagram of the model which can be detected by using the fidelity susceptibility.

\section{Discussion}
In the paper, we employ the entanglement entropy to study numerically the hard-core-boson model with nearest and next-nearest-neighbor interactions. By using exact diagonalization and the density-matrix renormalization group, the effects of interactions on the entanglement entropy in the model are investigated. It is found that the first derivative of the entanglement entropy can detect all the quantum phase transitions in the model, and the phase diagram is given. We also test the ability of the fidelity susceptibility to detect the phase transitions. It is found that the fidelity susceptibility can predict charge-density-wave II to bond order and charge-density-wave I to superfluid transitions, however, the fidelity susceptibility cannot predict the superfluid to bond order transition for finite-size systems.

\vspace{0.3cm}
\begin{acknowledgments}
J. R. thanks Wen-Long You and Yin-Zhong Wu for their helpful discussions. Financial supports from the National Natural Science Foundation of China (Grants No:11104021, 11174114, and 11274054) is gratefully acknowledged.
\end{acknowledgments}

\end{document}